\documentclass{iopart}
\usepackage{amsfonts,graphicx}
\def\vec#1{\bf #1}

\def\ket#1{| #1 \rangle}
\def\bra#1{\langle #1 |}
\def\ip#1#2{\langle #1 \mid #2 \rangle}

\def\norm#1{|| #1 ||}
\def\diag{\mbox{\rm diag}}
\def\dim{\mbox{\rm dim}}
\def\tr{\mbox{\rm Tr}}
\def\diag{\mbox{\rm diag}}
\def\uu{\mbox{\rm U}}
\def\su{\mbox{\rm SU}}
\def\sp{\mbox{\rm Sp}}
\font\liealg=eufm10
\def\dd{{\liealg D}}
\def\pp{{\liealg P}}

\def\O{{\cal O}}
\def\RR{\mathbb{R}}
\def\CC{\mathbb{C}}

\newtheorem{theorem}{Theorem} 
\newtheorem{proposition}{Proposition} 
\newtheorem{corollary}{Corollary} 
 
\newcounter{exNo}
\newenvironment{example}{\refstepcounter{exNo}\par\parskip1ex\noindent{\bf Example \arabic{exNo}:} \nobreak}{\par\noindent}
\newenvironment{proof}{\noindent{\sc Proof:} }{\rule{.6em}{0.6em}\par\noindent}

\newif\ifpdflatex\pdflatextrue
\makeatletter\@ifundefined{pdfoutput}{\pdflatexfalse}\makeatother
\def\myincludegraphics[#1]#2#3{%
\ifpdflatex \includegraphics[#1]{#2}
\else       \includegraphics[#1]{#3}
\fi}

\begin{document}
\title{Orbits of quantum states and geometry of Bloch vectors for $N$-level systems}
\author{S. G. Schirmer} 
\address{Department of Applied Maths and Theoretical Physics,
         University of Cambridge, Wilberforce Road, Cambridge, CB3 0WA, UK}
\address{Department of Engineering, Division F, Control Group, 
         University of Cambridge, Trumpington Street, CB2 1PZ, UK}
\author{T. Zhang} 
\address{Department of Mathematics and Statistics, Murray State University, 
         Murray, KY 42071, USA}
\author{J. V.  Leahy} 
\address{Department of Mathematics and Institute of Theoretical Science, 
        University of Oregon, Eugene, Oregon, 97403, USA}
\ead{\mailto{sgs29@cam.ac.uk},\mailto{tan.zhang@murraystate.edu},
     \mailto{leahy@math.uoregon.edu}}
\date{\today}
\begin{abstract}
Physical constraints such as positivity endow the set of quantum states with 
a rich geometry if the system dimension is greater than two.  To shed some 
light on the complicated structure of the set of quantum states, we consider 
a stratification with strata given by unitary orbit manifolds, which can be 
identified with flag manifolds.  The results are applied to study the geometry
of the coherence vector for $n$-level quantum systems.  It is shown that the 
unitary orbits can be naturally identified with spheres in $\mathbb{R}^{n^2-1}$ 
\emph{only} for $n=2$.  In higher dimensions the coherence vector only defines
a non-surjective embedding into a closed ball.  A detailed analysis of the 
three-level case is presented.  Finally, a refined stratification in terms 
of symplectic orbits is considered.
\end{abstract}
\pacs{03.65.-w,03.65.bz}
\maketitle

\section{Introduction}

The Bloch vector~\cite{Bloch} provides a representation of the quantum states of
a two-level system in terms of real observables, and allows the identification
of quantum states with points in a closed ball in 3D Euclidean space, the Bloch
ball, which has proven to be extremely useful.  In quantum information theory, 
for instance, the states of a single qubit can be identified with points on the 
surface of the Bloch ball (if the state is pure) or points inside the ball (if 
the state is mixed).  Unitary operations can be interpreted as rotations of this
ball, and dissipative processes as linear or affine contractions of the Bloch 
ball~\cite{Bloch,Ruskai}.  

Naturally, many efforts have been made to generalize the Bloch ball to higher
dimensions by defining a generalized coherence vector similar to the Bloch vector
for two-level systems.  However, while it is easy to define a Bloch~\cite{Hioe}
or general coherence vector~\cite{Lendi,Alicki} for $N$-level systems, it has 
become clear that the geometry of quantum states in higher dimensions is far more
complex than in the two-level case.  Some efforts have been made at determining 
the set of Bloch vectors corresponding to physical states for higher dimensional
systems using the higher trace invariants or Casimir invariants, for instance by 
Kimura~\cite{Kimura}, who also demonstrated the complicated and asymmetric 
structure of the set of Bloch vectors in higher dimensions.  Byrd and Khaneja 
\cite{Byrd} independently provided a similar characterization of the positivity
of the density matrix in terms of the coherence vector representation.

In this paper we pursue a different approach to study the structure of quantum 
states for higher dimensional systems, and the origin of the difference between 
the two-level case and higher dimensions.  In section II we define a natural
stratification of the set of density matrices in terms of unitary orbits.  We 
show that the unitary orbits can be identified with flag manifolds and determine
their dimensions.  For two-level systems this stratified set can be identified 
with a smooth real manifold with boundary, the 3D Bloch ball, with strata given
by concentric spheres.  

In section III we briefly review the definition of the Bloch vector and define 
a general coherence vector, which can be naturally embedded into a closed ball
in $\mathbb{R}^{n^2-1}$.  This embedding is surjective for two-level systems, 
hence allowing the identification of physical states with points in this closed
ball.  We show that for higher dimensional systems, however, the embedding is 
never surjective due to positivity constraints, and the dimensions of the orbit 
manifolds vary depending on the \emph{multiplicities} of the eigenvalues of the
states in each stratum.  Furthermore, for two-level systems there exists a total
ordering of the strata given by the length of the Bloch vector, or the distance
of the unitary orbit from the center of the Bloch ball -- the Bloch vector of 
pure states has length one, and the shorter the Bloch vector, the more mixed the
state is.  For higher dimensional systems we show that the length of the coherence
or generalized Bloch vector is no longer a sufficient measure for the disorder 
of the system.  A detailed analysis of the three-level case concludes section III.

Finally, in section IV we briefly consider a refined stratification of the set of
density matrices defined by the action of the symplectic group.  The symplectic 
group is of interest because it occurs naturally for quantum systems with certain 
dynamical symmetries (such as atomic systems with degenerate energy levels) and
it is the only proper subgroup of the unitary group that acts transitively on 
certain unitary orbits \cite{JPA35p4125}.  We show that the symplectic orbits 
of pseudo-pure states agree with the unitary orbits and provide bounds on the 
dimensions of other symplectic orbits, showing that the symplectic orbits 
generally have much lower dimension.  From a control point of view this means
that a $2n$-dimensional Hamiltonian control system with dynamical Lie group 
$\sp(n)$ is pure-state controllable but we cannot control generic ensembles.  

\section{Density matrices and unitary orbits}

Throughout this paper we restrict our attention to quantum systems whose Hilbert
space is a finite-dimensional complex vector space $\CC^{n}$, where $n$ is an 
integer greater than one, and for simplicity we will often use $V$ to denote 
$\CC^n$ with the standard Hermitian inner product $\langle \cdot,\cdot \rangle$.  
Any physical state of this system can be represented by a density operator, i.e.,
a positive semi-definite (self-adjoint) linear operator with trace one.  The 
subset of rank-one density operators corresponds to pure states of the system, 
all other density operators to mixed states.

In the following we denote the set of all positive semi-definite operators on 
$V$ by $\pp(V)$, the subset of all density matrices by $\dd(V)$, and the subset 
of pure states by $\dd_1(V)$.  We also define the class of pseudo-pure states or 
pure-state-like ensembles as the set of density operators $\dd_{1'}(V)$ whose 
spectrum consists of exactly two distinct eigenvalues with multiplicities one
and $n-1$, respectively.  It is well known that $\dd(V)$ forms a convex subset 
of the set of Hermitian matrices since given two density matrices $\rho_1,\rho_2 
\in\dd(V)$, the straight line path defined by $\Gamma(t):=(1-t)\rho_1+t\rho_2$ 
for $t\in [0,1]$ is contained in $\dd(V)$.  To see this, note that $\Gamma(t)$ 
is Hermitian, $\tr(\Gamma(t))=1$ and $\ip{ \Gamma(t)x}{x} =(1-t)\ip{\rho_1x}{x}
+t\ip{\rho_2x}{x}\ge 0$ for all $x\in\CC^n$ and $t\in [0,1]$.

We define the orbit of a quantum state $\rho$ under the action of the dynamical
Lie group $G$ to be the set $G\cdot\rho:=\{g\cdot \rho\cdot g^{-1}\ |\ g\in G\}$.
The orbits endow the set of quantum states with the structure of a stratified set.
In principle, any Lie group $G$ acting on the set of density operators defines a
stratification of $\dd(V)$.  However, the orbits under the action of $G=\uu(n)$ 
are of particular interest since they determine the most general evolution of the 
quantum states in a closed system.  From a control point of view, the unitary orbit
of a state represents the maximal set of states that are reachable from the given 
state via (open-loop) coherent control, or in the language of quantum computing, 
by applying a unitary gate to the state.  

Before we attempt to classify the orbits, we recall the following standard result
from linear algebra:
\begin{proposition}\label{prop:a}
Let $\rho_1$ and $\rho_2$ be two density matrices.  The following are equivalent:
\begin{enumerate}
\item $\rho_1$ and $\rho_2$ are unitarily equivalent, i.e., $\rho_2=U \rho_1 
      U^\dagger$ for some unitary matrix $U$.
\item $\rho_1$ and $\rho_2$ have the same spectrum, i.e., the same eigenvalues 
      including multiplicity.
\item $\tr(\rho_1^r)=\tr(\rho_2^r)$ for all $r=1,2,\ldots, n$.
\end{enumerate} 
\end{proposition}

This result shows immediately that the orbit of a density matrix under $\uu(n)$
is uniquely determined by its spectrum, i.e., two density matrices belong to the
same unitary orbit if and only if they have the same eigenvalues $\lambda_i$ with
the same multiplicities $n_i$.  Each orbit $\O$ can therefore be represented by a
canonical diagonal density matrix of the form
\begin{equation}
  \rho = \diag(\lambda_1 I_{n_1}, \ldots, \lambda_r I_{n_r}),
\end{equation}
where the eigenvalues shall be ordered such that $\lambda_i>\lambda_j$ for $i<j$
to ensure a unique representation.  Since the $\lambda_i$ can be arbitrary real 
numbers in $[0,1]$ provided they satisfy $\sum_{i=1}^r n_i \lambda_i =1$, we see 
immediately that there are infinitely many distinct orbits corresponding to the 
(uncountably) infinitely many possible choices for the $\lambda_i$.  Thus, we can
say that the unitary group $\uu(n)$ partitions the set of density matrices $\dd(V)$
into an uncountably infinite family of (distinct) orbits or strata.

We can define a (partial) ordering on this stratification via majorization.  Let
$\rho_1$, $\rho_2$ be two density operators with eigenvalues $a_m^{(i)}$, $i=1,2$,
counted with multiplicity and ordered in a nonincreasing sequence. $\rho_1\prec
\rho_2$ if 
\begin{equation}
  \begin{array}{ll}
  \sum_{m=1}^k a_m^{(1)} \le \sum_{m=1}^k a_m^{(2)}, &\quad k=1,\ldots,n-1\\  
  \sum_{m=1}^n a_m^{(1)} =   \sum_{m=1}^n a_m^{(2)},
  \end{array} 
\end{equation}
For instance, $\rho_1=\frac{1}{5}\diag(1,1,3)\prec\rho_2=\frac{1}{5}\diag(2,2,1)$ 
since $1\le 2$, $1+1\le 2+2$ and $1+1+3=2+2+1$.  Majorization has been shown to be
a useful way to compare the degree of disorder of physical systems~\cite{Nielson} 
and naturally defines a partial ordering on the unitary orbits (strata) via $\O_1
\prec\O_2$ if $\O_i=\O[\rho_i]$ and $\rho_1\prec\rho_2$.  However, note that only
some orbits can be compared that way.  Consider $\rho_1=\frac{1}{8}\diag(5,2,1)$
and $\rho_2=\frac{1}{8}\diag(4,4,0)$. We have $5>4$ but $5+2<4+4$.  Hence, neither
orbit majorizes the other.

To determine the nature of the strata given by the orbits, we define the 
\emph{isotropy subgroup} or stabilizer at $\rho$ as the subgroup $G_\rho$ of 
elements in $G$ that leave $\rho$ invariant, i.e., for which we have $g\cdot 
\rho\cdot g^{-1}=\rho$.   We shall show that the orbit of an element $\rho
\in\dd(V)$ under the unitary group $\uu(n)$ can be identified with a certain
type of manifold called a flag manifold.  For the purpose of proving this 
result we observe that $\uu(n)$ is a compact Lie group and hence a compact 
topological group%
\footnote{A topological group is a topological space $X$ endowed with a group
structure that allows us to ``multiply'' elements of the space and compute 
inverses such that both operations are continuous with respect to the topology.
The unitary group, for instance, is a multiplicative group since multiplication
of two unitary matrices gives a unitary matrix, every unitary matrix has an 
inverse given by the Hermitian conjugate, and the identity provides a neutral 
element.  Furthermore, as a subset of the complex matrices the unitary group is
naturally endowed with a topology that allows us to separate two unitary 
matrices by open sets, and matrix multiplication and Hermitian conjugation are 
continuous with respect to this topology.}
and the space of density matrices is a Hausdorff space%
\footnote{A Hausdorff space basically is a space endowed with a topology that
allows us to separate points by disjoint open sets.}, 
and we have the following result (see \cite{refGB}, for instance):

\begin{proposition} \label{prop:b}
If $G$ is a compact topological group acting on a Hausdorff space $X$ and 
$G_x$ is the isotropy group at $x$ then the map $\phi: G/G_x\mapsto G\cdot x$ 
is a homeomorphism.
\end{proposition}

\begin{theorem} \label{thm:one}
Let $\uu(n)$ act on $\dd(V)$ by conjugation and let $\rho$ be a quantum state
with $r\ge 1$ distinct eigenvalues $\lambda_i$ with (geometric) multiplicity 
$n_i$. Then the orbit of $\rho$ is homeomorphic to the flag manifold
\[
   \uu(n)/ [\uu(n_1)\times\uu(n_2)\times\cdots\times\uu(n_r)]
\]
of real dimension $n^2-\sum_{i=1}^r n_i^2$.
\end{theorem}

\begin{proof} 
Let $E_i$ be the eigenspaces of $\rho$ with $\dim E_i = n_i$.  Since $\rho$ 
is unitarily equivalent to the diagonal matrix $\diag(\lambda_1 I_{n_1}, 
\ldots, \lambda_r I_{n_r})$, we have an orthogonal direct sum decomposition 
of $V=\mathbb{C}^n$ of the form $V= E_1 \oplus \cdots \oplus E_r$.  
$g\in U(n)$ stabilizes $\rho$ if and only if $g$ preserves the eigenspaces
$E_i$, i.e., the restriction of $g$ to each eigenspace must be an isometry, 
i.e., $g$ preserves the eigenspaces $E_i$.  Hence, the orbit of $\rho$ is 
homeomorphic to the flag manifold $\uu(n)/[\uu(n_1)\times\uu(n_2)\times\cdots
\times\uu(n_r)]$ by Proposition \ref{prop:b}.
\end{proof}

\begin{corollary} \label{cor:one}
If $\rho \in \dd_{1'}(V)$ (pseudo-pure state) then the orbit of $\rho$ is 
homeomorphic to $\uu(n)/[\uu(1)\times \uu(n-1)]$, which is homeomorphic to 
the complex projective space $\mathbb{CP}^{n-1}$.
\end{corollary}

To illustrate the result, we explicitly compute the orbits under the action of 
$\uu(n)$ for $n=2$ and $n=3$.  

\begin{example} \label{ex:one}
Let $\rho$ be a $2\times 2$ density matrix.  $\rho$ is unitarily equivalent to
$\diag(r,1-r)$ with $0 \le r \le 1$.  If $r=1-r$ then $\rho=\frac{1}{2}I_2$ and 
the orbit of $\rho$ is homeomorphic to $\uu(2)/\uu(2)$, i.e., a single point.  
Otherwise, its orbit is homeomorphic to $\uu(2)/[\uu(1) \times \uu(1)] \simeq
\mathbb{CP}^1$.  

Since $\mathbb{CP}^1$ is diffeomorphic to the sphere $S^2$, this shows that any 
$\uu(2)$ orbit of a two-level system is homeomorphic to $S^2$, except the trivial
orbit of the completely random ensemble $\frac{1}{2} I_2$ which consists of a 
single point.  

Furthermore, note that the requirement of positivity of $\rho$ reduces to $0\le 
r \le 1$ and hence $r^2 \le r$, or equivalently, $\tr(\rho^2) = r^2+(1-r)^2 = 
1-2r+2r^2 \le 1-2r+2r =1$.  We shall see that this implies that the set of all 
$2\times 2$ density matrices (the union of all orbits) is homeomorphic to a 
closed ball in $\RR^3$.  
\end{example}

Thus, we have a neat mathematical justification for the Bloch ball description 
of a two-level system, which will be discussed in detail later.

\begin{example} \label{ex:two}
Let $\rho$ be a $3\times 3$ density matrix.  If $\rho$ has only one eigenvalue 
with multiplicity $3$ then $\rho=\frac{1}{3}I_3$ and its orbit is homeomorphic
to $\uu(3)/\uu(3)$, which is a single point as before.

If $\rho$ has two distinct eigenvalues then it is a pseudo-pure state unitarily
equivalent to $\rho=\diag(1-2a,a,a)$ where $0\le a\le 1$ and $a\neq\frac{1}{3}$.
Its isotropy subgroup is therefore $\uu(1)\times\uu(2)$, and its orbit is 
homeomorphic to $\uu(3)/[\uu(1)\times\uu(2)]$ and has dimension $9-1-4=4$.

If $\rho$ is a generic ensemble with three distinct eigenvalues $a,b,c$ then 
its canonical form is $\rho=\diag(a,b,c)$ and its isotropy subgroup is $\uu(1)
\times \uu(1)\times \uu(1)$.  Hence, its orbit is homeomorphic to $\uu(3)/[\uu(1)
\times \uu(1) \times \uu(1)]$ and has real dimension $9-3=6$. 
\end{example}

The results of the previous example are summarized in the Table~\ref{table:one}.
The table also provides a complete classification of the orbits for $n=4$.  
\begin{table}
\[\begin{array}{|l|l|l|}
\hline
n=3                 & \mbox{manifold}                  & \mbox{dim.}  \\ \hline
\rho = \diag(a,a,a) & point                            & 0 \\
\rho = \diag(a,b,b) & U(3)/[S^1 \times U(2)]           & 4 \\
\rho = \diag(a,b,c) & U(3)/[S^1 \times S^1 \times S^1] & 6 \\
\hline
\end{array}\]

\[\begin{array}{|l|l|l|}
\hline
n=4                   & \mbox{manifold}               & \mbox{dim.}  \\ \hline
\rho = \diag(a,a,a,a) & point                                       & 0 \\
\rho = \diag(a,b,b,b) & U(4)/[S^1 \times U(3)]                      & 6 \\
\rho = \diag(a,a,b,b) & U(4)/[U(2) \times U(2)]                     & 8 \\
\rho = \diag(a,b,c,c) & U(4)/[S^1 \times S^1 \times U(2)]           & 10 \\
\rho = \diag(a,b,c,d) & U(4)/[S^1 \times S^1 \times S^1 \times S^1] & 12 \\
\hline
\end{array}\]
\caption{Manifolds and their dimension for the unitary orbits of quantum states 
based on their canonical form.  All parameters $a,b,c,\ldots$ in the table above
are in $[0,1]$ such that $\tr(\rho)=1$ and different letters represent different
values.} \label{table:one}
\end{table}

The previous two examples clearly show the difference between two-level systems
and higher dimensional systems ($n>2$).  While all orbit manifolds (except the
trivial orbit of the completely random ensemble) for two-level systems are 
homemorphic to a sphere, no such homeomorphism is possible in the latter case, 
i.e., the orbit manifolds for higher dimensional systems can \emph{never} be 
\emph{identified} with spheres in a higher dimensional Euclidean space.  That 
is, although we can always \emph{embed} the quantum states of the system in a 
compact subset (closed ball) of a real vector space of sufficiently high dimension, 
there is no one-to-one correspondence between spheres in this ball and orbits of
quantum states, \emph{except for $n=2$}, which explains the difficulties one 
encounters when trying to generalize intuitive reasoning valid for the Bloch 
ball for $n=2$ to Bloch vectors in higher dimensions.

\section{Coherence vector and embeddings of quantum states}

\subsection{Definition of Bloch or coherence vector}

For a two-level system any density operator can be expanded as
\[
   \rho = \frac{1}{2} (I_2 + x \sigma_x + y \sigma_y + z \sigma_z)
\]
where $I_2$ is the 2D identity matrix, and 
$\sigma_x=\ket{1}\bra{2}-\ket{2}\bra{1}$, 
$\sigma_y=\rmi(\ket{1}\bra{2}+\ket{2}\bra{1})$ and 
$\sigma_z=\rmi(\ket{1}\bra{1}-\ket{2}\bra{2})$ 
are the usual (unnormalized) 2D Pauli matrices.  The coordinates $x$, $y$ and $z$
are real since the Pauli matrices are Hermitian and $*=\tr(\rho\sigma_*)$ for $*=
x,y,z$.  Hence, the state of any two-level system can be characterized completely
by the real vector $\vec{s}=(x,y,z)$, called the Bloch vector.

For an $n$-level system we can proceed in a similar fashion.  Let 
\begin{equation}\begin{array}{l}
  \sigma_{r,s}^x = \ket{r}\bra{s} - \ket{s}\bra{r}, \quad 
  \sigma_{r,s}^y = \rmi(\ket{r}\bra{s} + \ket{s}\bra{r}), \\
  \sigma_{r}^z   = \rmi \sqrt{\frac{2}{r+r^2}} \left(\sum_{k=1}^r \ket{k}\bra{k} 
                   - r\ket{r+1}\bra{r+1} \right)
\end{array}
\end{equation}
for $1 \le r \le n-1$ and $r<s\le n$ be the generalized Pauli matrices in 
dimension $n$.  The set $\{\tilde{\sigma}_k\}_{k=1}^{n^2-1}=\{\sigma_{r,s}^x, 
\sigma_{r,s}^y, \sigma_{r}^z \, |\, 1\le r <n,\, r<s\le n\}$ forms a basis
for the space of $n \times n$ traceless Hermitian matrices satisfying the 
\emph{orthogonality} condition
\begin{equation}
   \ip{\tilde{\sigma}_k}{\tilde{\sigma}_\ell} 
 = \tr(\tilde{\sigma}_k \tilde{\sigma}_\ell) 
 = 2\delta_{k,\ell}.
\end{equation}
Every density matrix can be expanded with respect to this basis
\begin{equation} \label{eq:bloch}
  \rho = \frac{1}{n} I_n + 
         \frac{1}{2} \sum_{k=1}^{n^2-1} \tilde{s}_k \tilde{\sigma}_k
\end{equation}
where $\tilde{s}_k = \tr(\rho\tilde{\sigma}_k)$ for $k=1,\ldots,n^2-1$.  The 
resulting real vector $\tilde{\vec{s}}=(\tilde{s}_k)_{k=1}^{n^2-1}$ is the 
Bloch vector of the $n$-dimensional system.

Although the Bloch vector defined above is useful, it is generally more elegant, 
and often more convenient, to work with an \emph{orthonormal} basis.  To this 
end, we define the normalized Pauli matrices $\{\sigma_k\}_{k=1}^{n^2-1}=
\{\frac{1}{\sqrt{2}}\tilde{\sigma}_k\}_{k=1}^{n^2-1}$, which satisfy the 
\emph{orthonormality} condition
\begin{equation}
  \ip{\sigma_k}{\sigma_\ell} = \tr(\sigma_k \sigma_\ell) = \delta_{k,\ell}.
\end{equation}
Furthermore, $\{\sigma_k\, |\, k=1,\ldots, n^2-1\}$ together with $\sigma_0=
\frac{1}{\sqrt{n}}I_n$ forms an orthonormal basis for all Hermitian $n \times n$
matrices, and we can expand any density matrix in terms of this ON basis
\begin{equation} \label{eq:coherence}
  \rho = \sum_{k=0}^{n^2-1} s_k \sigma_k
\end{equation}
where $s_k=\tr(\rho \sigma_k)$ for $k = 0,\ldots, n^2-1$.  Since $1=\tr(\rho)=
\sqrt{n}s_0$, we have $s_0=\frac{1}{\sqrt{n}}$ for all density matrices.  Hence,
$\rho$ is completely determined by the real $n^2-1$ vector $\vec{s}=(s_1,\ldots,
s_{n^2-1})$.  This vector is often called the general coherence vector.  Equations
(\ref{eq:bloch}) and (\ref{eq:coherence}) are equivalent, and we easily see that 
$\tilde{\vec{s}}=2\vec{s}$, i.e., the standard Bloch vector differs from the 
coherence vector only by a factor of $2$.  


It is easy to verify that the mapping that takes $\rho$ to the real coherence 
vector $\vec{s}$, or equivalently the Bloch vector $\tilde{s}=2\vec{s}$, defines
an embedding of the density matrices into a closed ball in $\mathbb{R}^{n^2-1}$ 
for all $n>1$.  However, the two-level case is special in that the embedding
defined is surjective.  We shall now discuss the nature of the resulting 
differences between the $n=2$ and $n>2$ case, and provide a detailed analysis 
of the three-level case. 

\subsection{Bloch ball picture for $n=2$}

For two-level systems the embedding defined above is not only one-to-one but also
surjective, and hence provides a homeomorphism between orbits of density matrices
under $\uu(n)$ and the closed ball of radius one (Bloch vector as defined above) 
or radius $\frac{1}{2}$ (coherence vector as defined above) in $\mathbb{R}^3$.  
Unitary transformations of a density matrix can be interpreted as real rotations 
of the this ball.  Example \ref{ex:one} shows that the action of the unitary group
on the set of quantum states partitions it into an uncountably infinite number of 
distinct orbit manifolds, homeomorphic to concentric, two-dimensional spheres in
the Bloch ball, with the exception of the trivial orbit of the completely random 
ensemble, which is mapped to the single point at the center of the ball.  

It is also easy to see that the distance of an orbit from the center of the ball
is determined by $\tr(\rho^2)$ via $\tilde{r}=2\tr(\rho^2)-1$ (Bloch vector) or 
$r=\tr(\rho^2)-\frac{1}{2}$ (coherence vector).  Pure states ($\tr(\rho^2)=1$) 
have maximal distance from the center of the ball and hence form its boundary. 
Furthermore, the disorder of an orbit is completely determined by its distance
$r(\O)$ from the center, i.e., $\O_1 \prec \O_2$ if $r(\O_1) < r(\O_2)$.

From the point of view of controllability of quantum systems, it is also worth 
noting that all orbits (with the exception of the completely random ensemble) 
have the same dimension and geometry.  Hence, any group that acts transitively
on the class of pure states, for instance, will also act transitively on all 
classes of mixed states and vice versa.  Of course, the only such groups are 
$\uu(2)$ or $\su(2)$.  Hence, pure-state and mixed-state controllability are 
equivalent notions for two-level systems.

\subsection{Bloch ball picture for $n>2$}

For $n>2$ the situation is quite different due to the fact that the embedding
into a closed ball in $\RR^{n^2-1}$ defined by $\rho \mapsto \vec{s}$ is not 
surjective.  It is easy to see that the distance of each unitary orbit from the
center in $\mathbb{R}^{n^2-1}$ remains completely determined by $\tr(\rho^2)$:
\begin{equation}
  \norm{\vec{s}}^2 = \sum_{k=1}^{n^2-1} s_k^2 = \tr(\rho^2) - \frac{1}{n}.
\end{equation}
However, a glance at the dimensions of the orbits in Table \ref{table:one} shows
immediately that each orbit is only a submanifold of a sphere of a fixed distance
from the origin.  For instance, as we have shown in the previous section, the 
orbit of pure states for $n=3$ corresponds to a four-dimensional submanifold of 
the seven-dimensional boundary sphere with radius $\sqrt{1-\frac{1}{3}}$ in 
$\mathbb{R}^8$.  Since there is only a single orbit of pure states, the remainder
of the points on the boundary sphere do not correspond to physical states.  

For mixed states the situation is more complicated since each sphere of fixed 
radius $r$ from the origin now generally contains an uncountably infinite number
of distinct orbits, all satisfying $\tr(\rho_m)=1$ and $\tr(\rho_m^2)=r^2$ but 
differing in higher trace invariants, e.g., $\tr(\rho_m^k)\neq\tr(\rho_{m'}^k)$ 
for some $2<k\le n$.  The dimensions of the orbits contained within each sphere
vary depending on the type of ensemble but each sphere (except for the boundary) 
generally contains a set of positive measure of orbits corresponding to physical
states, and often a positive-measure set of points that do not belong to physical
orbits.  Moreover, the degree of disorder of an orbit can no longer be properly
characterized by its distance from the center of the ball.  Orbits contained in
the same sphere can often not be compared with respect to our partial ordering 
and may have different von-Neumann entropy.  Furthermore, orbits at different 
distances from the origin may have the same von-Neumann entropy.

\subsection{Analysis of $n=3$ case}

\begin{figure}
\begin{center}
\myincludegraphics[width=0.45\textwidth]{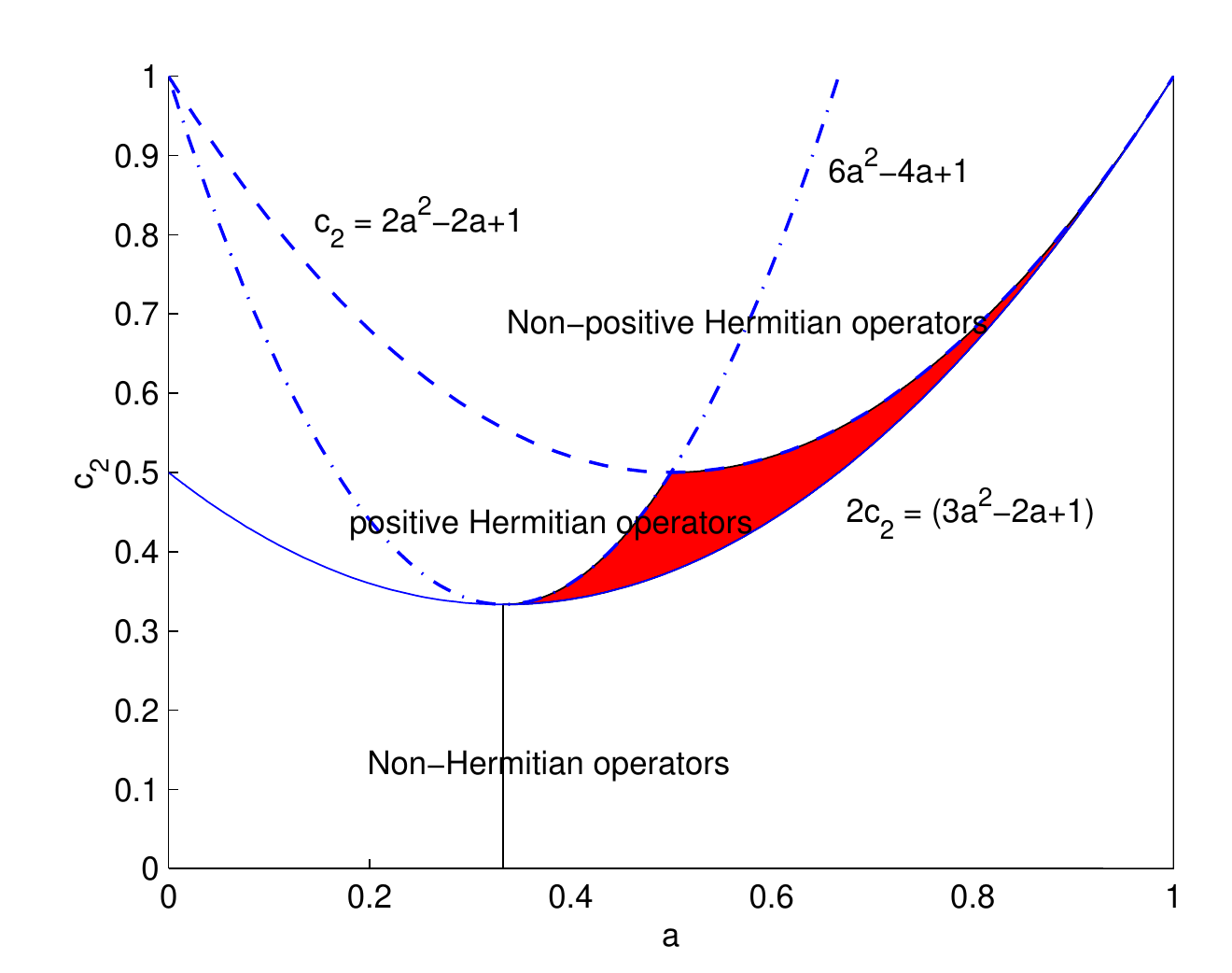}{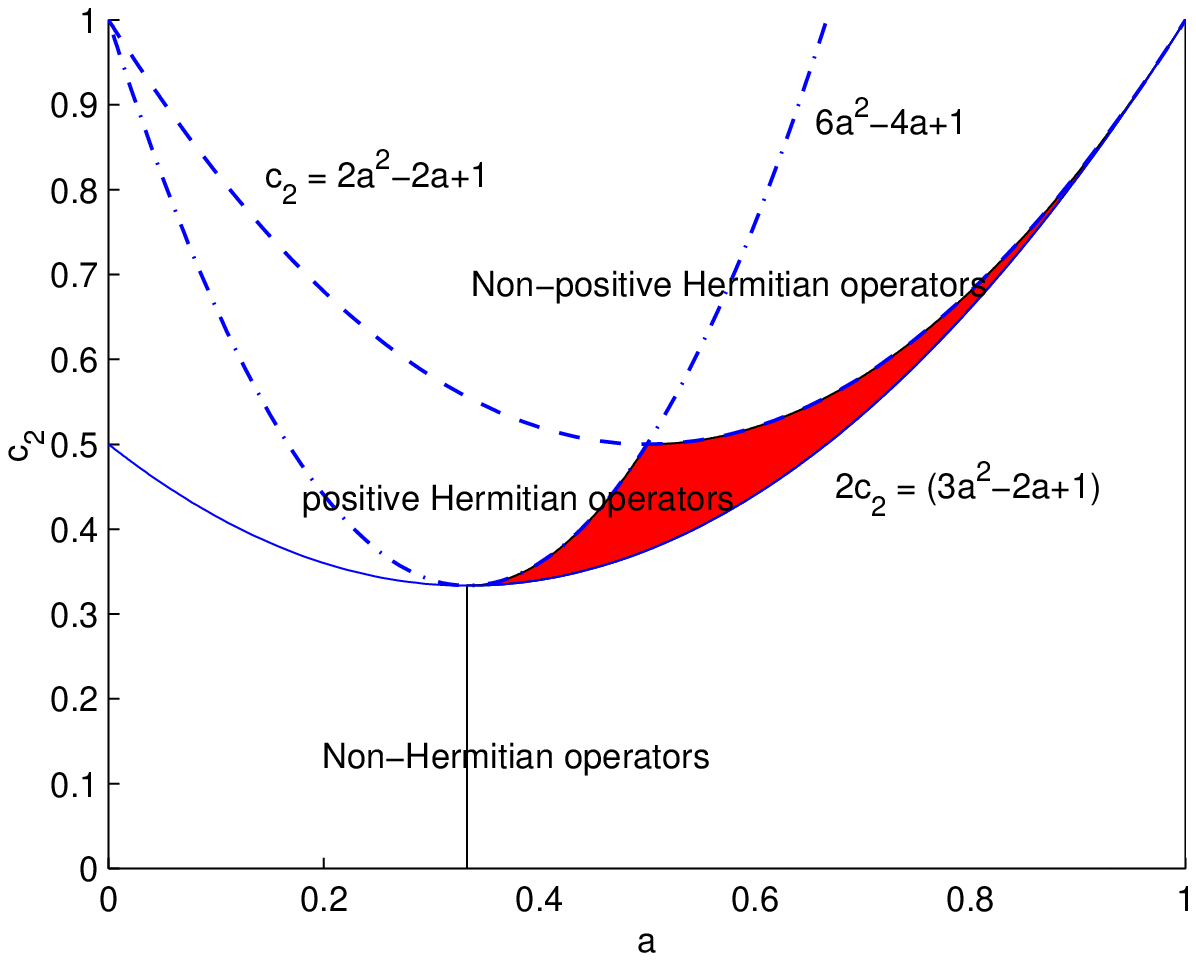}
\caption{Diagram indicating parameter values corresponding to physical orbits}\label{fig:one}
\end{center}
\end{figure}

To illustrate the general statements above, let us consider the $n=3$ case and 
the set of orbits determined by the family of states $\rho=\diag(a,b,c)$ with
$b,c$ given by
\begin{eqnarray*}
  b &=& \frac{1}{2}(1-a + K), \\
  c &=& \frac{1}{2}(1-a - K), \\
  K &=& \sqrt{-1+2a-3a^2+2c_2}.
\end{eqnarray*}
Note that we have $\tr(\rho)=a+b+c=1$ and $\tr(\rho^2)=c_2$ for all $a$.  However, 
for $\rho$ to represent a physical state $a,b,c$ must be real and have values in 
$[0,1]$.  These constraints imply that the argument of $K$ must be non-negative 
and $K\le 1-a$, or equivalently $K^2\le (1-a)^2$.  This yields the inequalities
\begin{eqnarray}
  3a^2-2a+1 &\le& 2c_2, \label{eqa} \\
  2a^2-2a+1 &\ge&  c_2, \label{eqb}
\end{eqnarray}
which must be simultaneously satisfied.  To ensure that there is a one-to-one
correspondence between parameter values $(a,c_2)$ and orbits, we further require
$a\ge b\ge c$.  Since $a+b+c=1$ is implies $a\ge \frac{1}{3}$.  The constraint
$b\ge c$ is automatically satisfied because $K$ is real and $\ge 0$, whereas the 
constraint $a\ge b$ implies:
\begin{equation} \label{eqc}
  6a^2-4a+ 1 \ge c_2.
\end{equation}
Inequality (\ref{eqa}) is satisfied for 
\[
  a \in \left[ \frac{1-K_1}{3}, \frac{1+K_1}{3} \right]
\]
for $K_1=\sqrt{6c_2-2}$.  Inequality (\ref{eqc}) is satisfied for 
\[
  a \in \left[ 0, \frac{1-K_1/6}{3} \right] \cup \left[\frac{1+K_1/6}{3},1\right].
\]
For $c_2\le\frac{1}{2}$ inequality (\ref{eqb}) is satisfied for $a\in [0,1]$, 
and $c_2>\frac{1}{2}$ it is satisfied for
\[
 a \in \left[0, \frac{1-K_2}{2} \right] \cup \left[\frac{1+K_2}{2},1 \right]
\]
for $K_2=\sqrt{2c_2-1}$.  Combining these inequalities and noting that $a\ge 
\frac{1}{3}$ leads to
\begin{eqnarray}
  c_2\le \frac{1}{2}: && a \in \left[ \frac{1}{3}(1+K_1/6),\frac{1}{3}(1+K_1)\right] \\
  c_2 > \frac{1}{2}:  && a \in \left[ \frac{1}{2}(1+K_2),\frac{1}{3}(1+K_1) \right] .
\end{eqnarray}
Fig.~\ref{fig:one} illustrates the situation. The solid curve corresponds to 
$3a^2-2a+1=2c_2$, the dash-dot line to $6a^2-4a+1=c_2$ and the dashed line to
$2a^2-2a+1=c_2$.  The points $(a,c_2)$ on the solid line correspond to orbits
of pseudo-pure states $\rho=\diag(a,\frac{1}{2}(1-a),\frac{1}{2}(1-a))$, and
the points on dash-dot line correspond to orbits of pseudo-pure states of the 
form $\rho=\diag(a,a,1-2a)$.  The points below the solid curve correspond to 
non-Hermitian matrices.  The points $(a,c_2)$ above the dashed line correspond 
to non-positive Hermitian matrices.  The points between these two curves 
represent physical states.  However, only points in the shaded region between
the curves satisfy all inequalities and represent unique orbits.

The figure also shows that for any $c_2\in (\frac{1}{3},1)$, there exists a 
positive-measure set (interval) of $a$-values that correspond to distinct physical
orbits, embedded in a sphere of fixed radius $r=\sqrt{c_2-\frac{1}{3}}$ from the 
origin.  This shows that the number of distinct orbits contained within each sphere
of radius $0<r<\sqrt{\frac{2}{3}}$ is uncountably infinite.  A unique orbit of 
fixed distance from the center of the ball exists only for the special cases $c_2=
\frac{1}{3}$ and $c_2=1$, the former corresponding to the trivial (i.e., zero-dimensional)
orbit of the completely random ensemble $a=b=c=\frac{1}{3}$ that forms the center 
of the ball, and the latter to the four-dimensional orbit of pure states contained
within the boundary sphere of the ball.  Most of these orbits are generic and hence
have dimension six.  However, each sphere also contains at least one (two if $c_2\le
\frac{1}{2}$) pseudo-pure orbits of dimension four, corresponding to the points 
$(a,c_2)$ on the boundary curves $2c_2=3a^2-2+1$ and $c_2=6a^2-4+1$, respectively.

Furthermore, for $c_2\le\frac{1}{2}$ all values of $(a,c_2)$ that correspond to 
Hermitian matrices actually correspond to positive Hermitian operators, i.e., 
physical states.  Hence, the union of all orbits with $c_2\le\frac{1}{2}$ fills 
a ball of radius $\frac{1}{\sqrt{6}}$.  For $c_2 \ge \frac{1}{2}$, however, the 
positivity constraint kicks in and eliminates more and more $a$-values as $c_2$ 
approaches one.  This means that the physical orbits inside the ball get sparser
as $c_2$ increases.  However, for each $c_2<1$ the orbits always occupy a positive
measure set of the sphere $S^7$ they are embedded in since for each $c_2<1$ there
is a 1D set of positive measure of $a$ values representing disjoint physical orbits
of dimension six.  Hence, the union of these orbits (not counting the one or two 
pseudo-pure orbits) occupies a set of positive measure inside each seven-dimensional
sphere $S^7$.  For $c_2=1$ the positivity constraint eliminates all $a$-values 
but $a=1$ and hence the boundary sphere contains only the single four-dimensional
orbit of proper pure states.

\begin{figure}
\begin{center}
\myincludegraphics[width=0.45\textwidth]{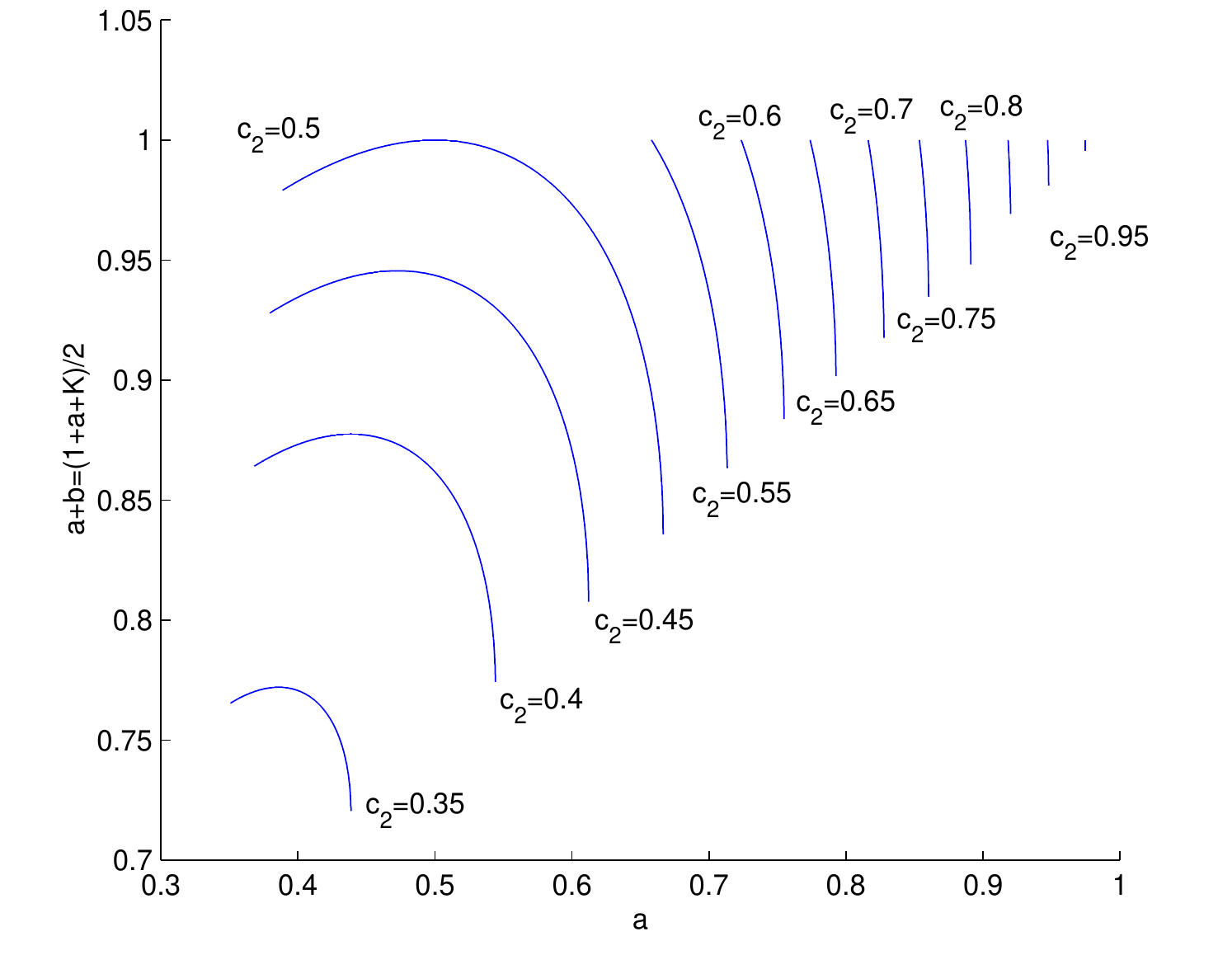}{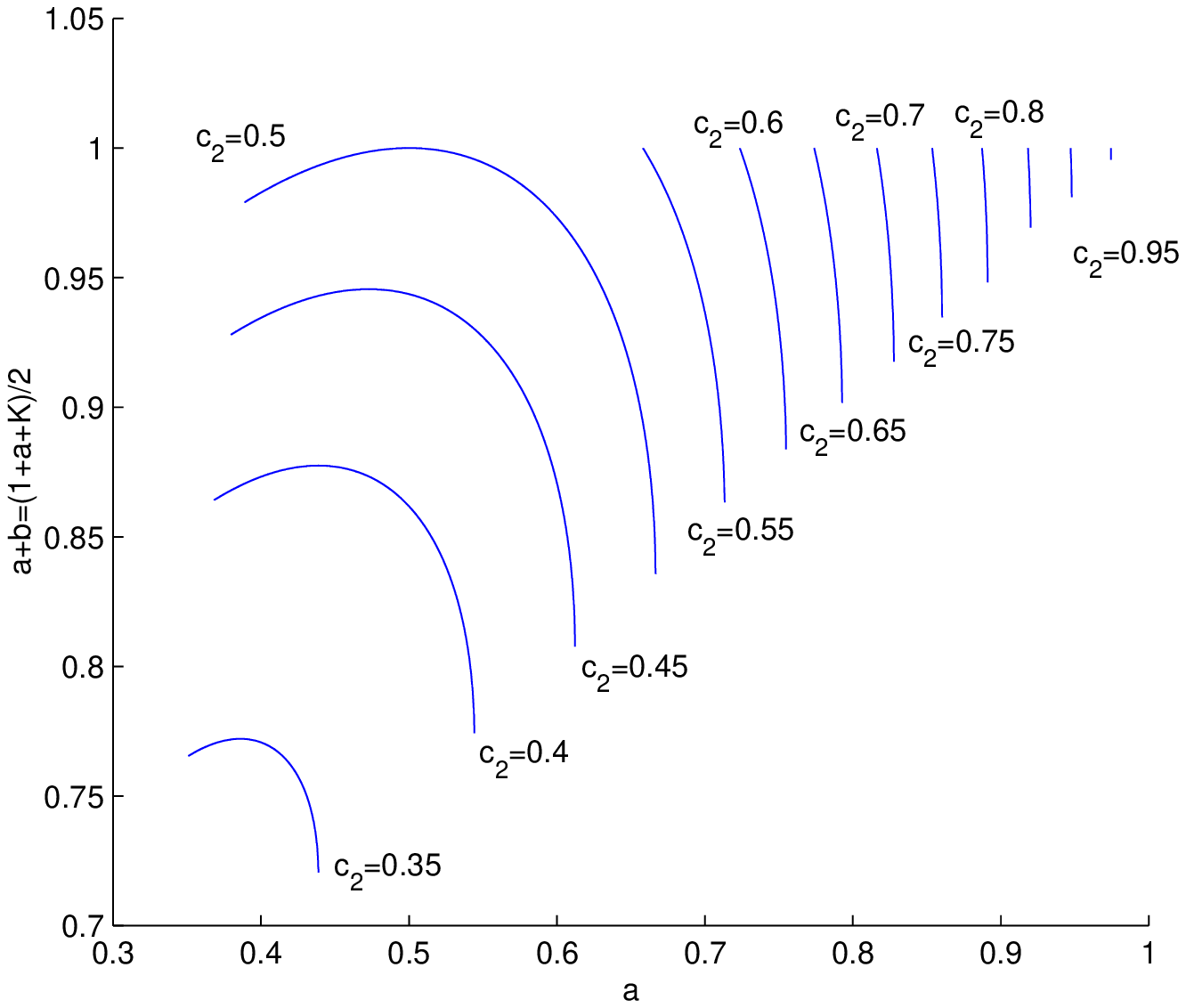}
\caption{$a+b=(1+a+K)/2$ as a function of $a$ for various values of $c_2$}\label{fig:two}
\end{center}
\end{figure}

\begin{figure}
\begin{center}
\myincludegraphics[width=0.45\textwidth]{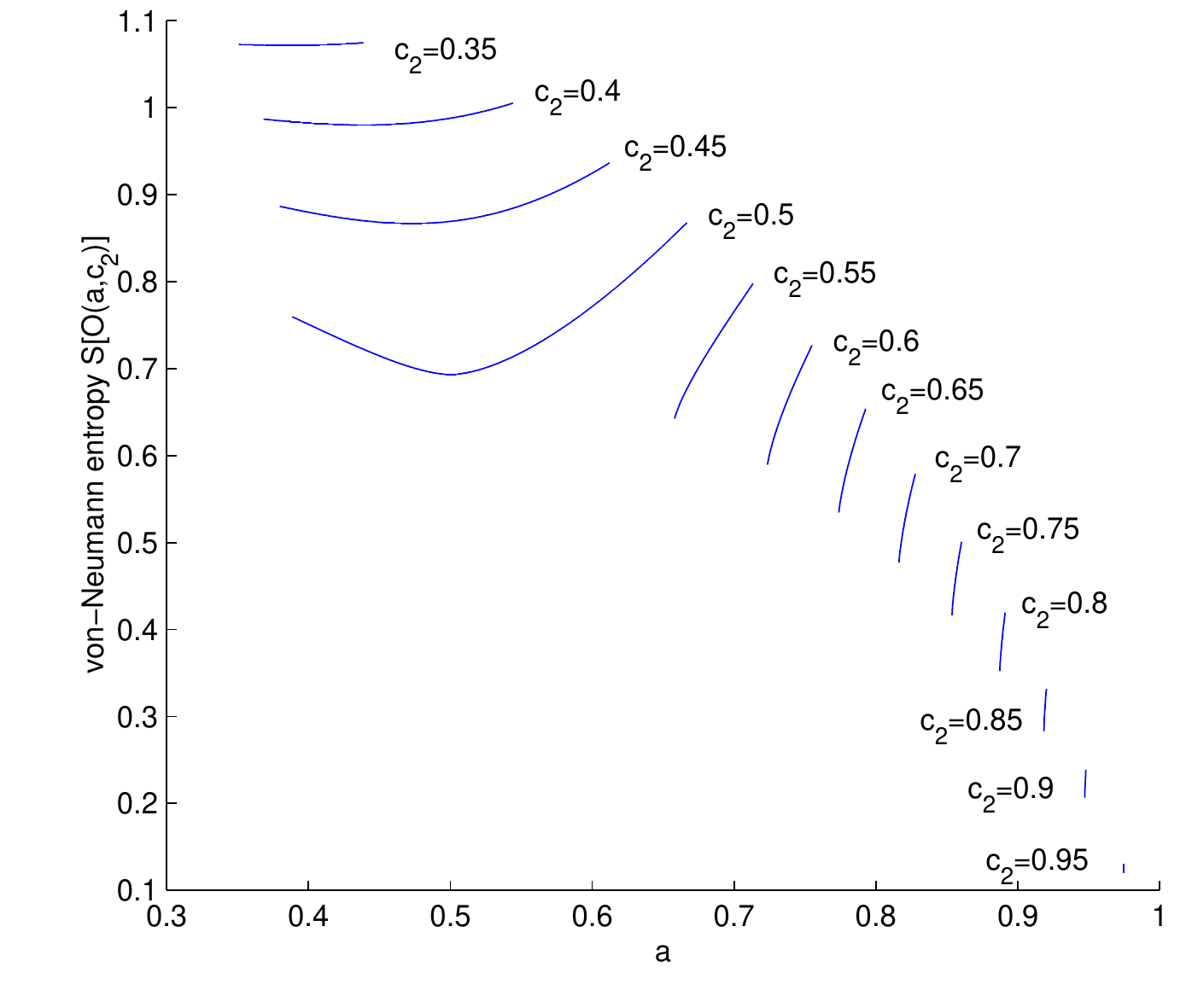}{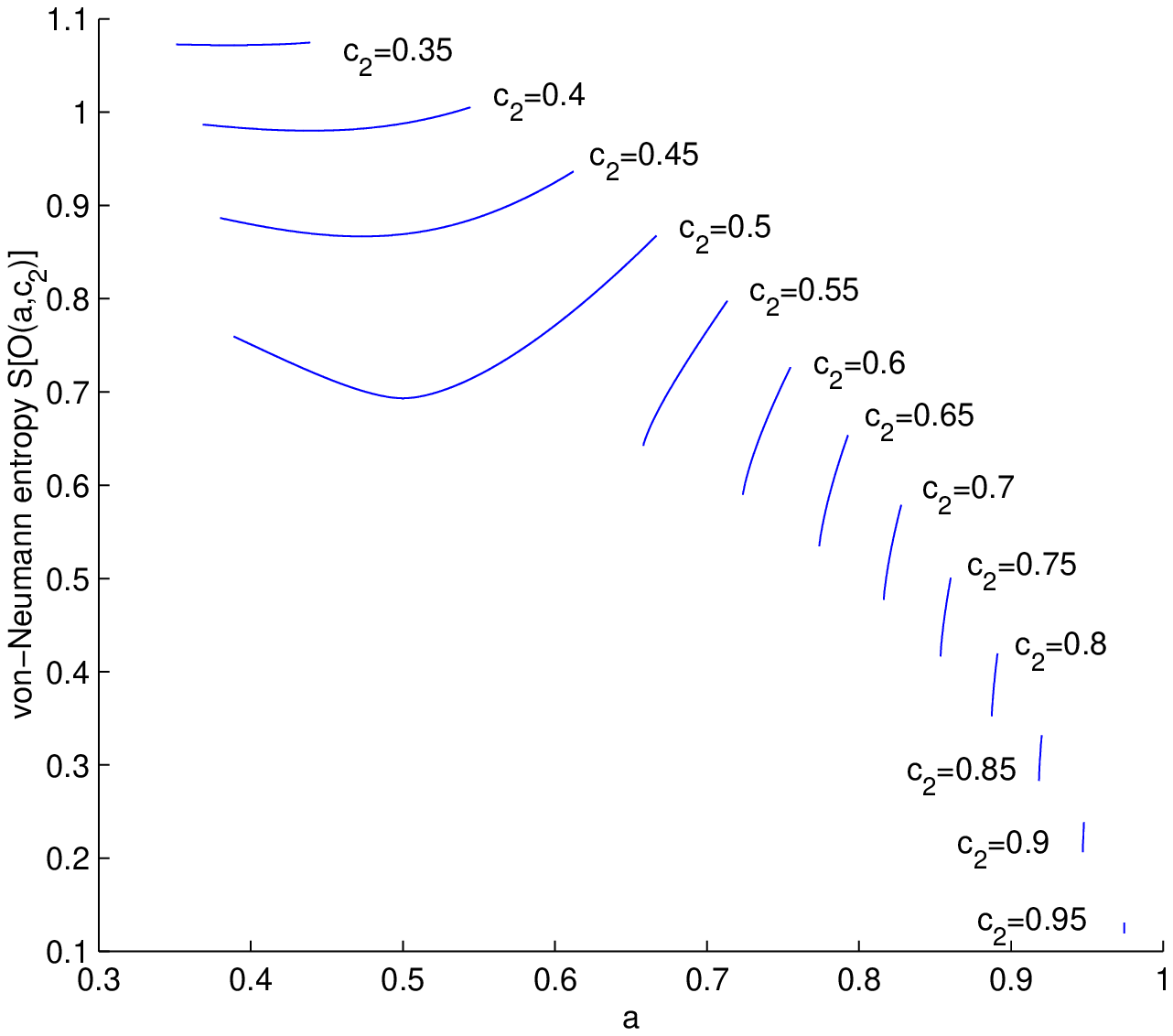}
\caption{von-Neumann entropy of orbits as a function of $a$ for various values of $c_2$}\label{fig:three}
\end{center}
\end{figure}

As regards the ordering of orbits, we have $\O[a,c_2] \prec \O[a,c_2']$ if $c_2<
c_2'$ since $K(a,c_2) < K(a,c_2')$ for $c_2 < c_2'$ and $a$ fixed.  In general, 
however, we cannot compare orbits that have the same distance from the origin 
since $a+b=\frac{1}{2}(1+a+K)$ is monotonically \emph{decreasing} over the valid
range of $a$ for $c_2>\frac{1}{2}$, and non-monotonic for $c\le\frac{1}{2}$, i.e., 
$a<a'$ typically implies $a+b>a'+b'$.  See Fig.~\ref{fig:two} for a plot of $a+b=
\frac{1}{2}(1+a+K)$ as a function of $a$ for various values of $c_2$.  
Fig.~\ref{fig:three}, for comparison, shows a plot of the von-Neumann entropy of
the orbits $S[\O] = -[a\log(a)+b\log(b)+c\log(c)]$ as a function of the largest
eigenvalue $a$ for various values of $c_2$.  Note that for $c_2\ge\frac{1}{2}$
fixed, the von-Neumann entropy increases as function of $a$, for $c<\frac{1}{2}$
it is non-monotonic as function of $a$.  Also note that for sufficiently large
values of $S$, there are many orbits of varying distance from the center with
the same von-Neumann entropy $S$ although for \emph{fixed} $a$ the orbits with
the largest distance from the center have the smallest entropy, which is in 
accord with our partial ordering of the orbits.

\section{Actions of the symplectic group}

In the previous sections we have shown that the action of the unitary group on
the set of quantum states endows it with the structure of a stratified set and
studied the properties of the strata defined by the unitary orbit manifolds.
This stratification of the set of density matrices was justified on physical
grounds since the dynamics of a (closed) quantum system is usually determined 
by the action of the unitary group.
 
Some physical systems, however, exhibit symmetries that restrict the dynamics
to a proper subgroup of the unitary group.  For instance, the dynamics of an
atomic system comprising two $n$-fold degenerate energy levels subject to 
coherent control fields of various polarizations is governed by the symplectic
group $\sp(n)$ due to dynamical symmetries \cite{JPA35p4125}.  Physical systems
that exhibit symplectic symmetry are also of special interest in quantum control
since they are pure-state controllable but lack mixed-state controllability 
\cite{JPA35p4125}.  This means, for instance, that we can steer the system 
from any pure initial state to any other pure state; however, if the system 
is initially in a mixed state, then it may not be possible to steer it to 
another mixed state even if this state is unitarily equivalent to the initial 
state, since the set of unitary operations at our disposal is limited and 
there may not be a sympletic unitary transformation that achieves the desired
aim \cite{JPA35p8551}.

The action of the symplectic group also induces a stratification of the set of 
density matrices.  Since the sympletic group $\sp(n)$ is a subgroup of the
unitary group $\uu(2n)$, it naturally follows that the sympletic orbits are
smaller than the unitary orbits.  Hence, there will be more sympletic orbits.
We can also think of the action of the symplectic group as partitioning the
unitary orbits into symplectic suborbits.  The stratification induced by the
symplectic group is therefore a refinement of the stratification induced by
$\uu(2n)$.  The remainder of this section is devoted to exploring the relation
between the symplectic and unitary orbits / stratification.

Mathematically, the symplectic group $\sp(n)$ is the subgroup of unitary 
transformations $A \in\uu(2n)$ that satisfy $A^T J A = J$ for 
\begin{equation} \label{eq:J}
   J= \left(\begin{array}{cc} 0 & I_n \\ - I_n & 0 \end{array} \right)
\end{equation}
where $I_n$ is the identity matrix in dimension $n$.  Note that, technically,
any group $G$ that satisfies $\{S^T J S = J \, | \, \forall S\in G\}$ for a 
matrix $J$ unitarily equivalent to the $J$ in (\ref{eq:J}) is a representation
of $\sp(n)$ but we shall assume the standard representation with $J$ as in 
(\ref{eq:J}) here.  The condition $S^T J S=J$ then implies that any $2n \times
2n$ complex matrix $S\in\sp(n)$ must be of the form
\begin{equation} \label{eq:sp-form}
  S = \left(\begin{array}{cc} 
                   A    & B \\
                   -B^* &  A^* 
            \end{array} \right)
\end{equation}
where $A$ and $B$ are $n \times n$ complex matrices and $A^*$ denotes the complex 
conjugate of the $A$.

We now show that $\sp(n)$ acts transitively on the unitary orbits of some states, 
but for the majority of states the symplectic orbits have lower dimension than the
unitary orbits.  The following results are an extension of earlier results showing
that $\sp(n)$ acts transitively only on pure-state-like and completely random 
ensembles \cite{JPA35p8551}.

\begin{proposition} \label{prop:sp:one}
If $\rho$ is the completely random ensemble $\frac{1}{2n}I_{2n}$ or a pseudo-pure
state then its orbit under $\sp(n)$ is the same as the orbit under $\uu(2n)$, i.e.,
$\sp(n)$ acts transitively on this orbit.
\end{proposition}

\begin{proof}
The orbit of any pseudo-pure state $\rho\in\dd_{1'}(V)$ under $\uu(2n)$ is 
homeomorphic to $\mathbb{CP}^{2n-1}$ by corollary \ref{cor:one}.  The assertion 
that $\sp(n)$ acts transitively on the unitary orbits of pseudo-pure states hence
follows directly from the well-known fact that $\sp(n)$ acts transitively on 
$\mathbb{CP}^{2n-1}$ via the isomorphism $\phi:\mathbb{H}^n \buildrel\cong\over
\rightarrow \mathbb{C}^{2n}$ discussed in \ref{appendix:A}.  Since the 
orbit of $\rho=\frac{1}{2n}I_{2n}$ under $\uu(2n)$ consists of a single point, 
the assertion that $\sp(n)$ acts transitively on this orbit is trivial.
\end{proof}

\begin{proposition} \label{prop:sp:two}
Let $\rho=\diag(\lambda_1,\cdots,\lambda_{2n})$ where $0\le\lambda_i\le 1$ are
the eigenvalues of $\rho$ counted with multiplicity, satisfying $\sum_{i=1}^{2n}
\lambda_i=1$. Then the orbit of $\rho$ is a homogeneous manifold of real dimension
at most $2n^2$.
\end{proposition}

\begin{proof} 
Since $\rho$ is diagonal the isotropy subgroup $G_\rho$ contains the maximal 
torus $T^{n}$ of $\sp(n)$, namely all matrices of the form $\diag(z_1,\cdots, 
z_n,z_1,\cdots, z_n)$ with $z_i\in\uu(1)\simeq S^1$.  Since $G_\rho$ is a 
closed subgroup of the Lie group $\sp(n)$, Proposition \ref{prop:b} implies
that the orbit is a homogeneous manifold of real dimension no more than
$\dim_{\mathbb{R}}(\sp(n)/T^{n})=n(2n+1)-n=2n^2$.
\end{proof}

For certain special cases we can improve this bound on the orbit dimensions.
\begin{proposition} \label{prop:sp:three}
If $\rho=\diag(\sigma_n,\sigma_n)$ where $\sigma_n$ is a diagonal $n\times n$ 
density matrix but not a multiple of $I_n$ then the sympletic orbit of $\rho$ 
is a homogeneous manifold of real dimension at most $2n^2-1$.  If $\rho=
\diag(aI_n, bI_n)$ with $0\le a, b\le 1$ and $a\neq b$ then the orbit of $\rho$
under $\sp(n)$ is a homogeneous manifold of real dimension $n^2+n$.
\end{proposition}

\begin{proof}
In the former case, observe that the isotropy subgroup at $\rho$ contains not only 
the maximal torus $T^{n}$, but all matrices of the form $\{zJ\ |\ z\in S^1\}$. Since
the matrix $J$ is not symplectic-equivalent to any element of the maximal torus 
$T^{n}$, the first statement follows immediately from the previous proposition.

In the second case, note that any element $Q$ that belongs to the isotropy subgroup
at $\rho$ must satisfy $Q\rho Q^\dagger=\rho$.  Since $Q$ must also be in $\sp(n)$,
it must be of the form (\ref{eq:sp-form}) for some $A,B\in\mbox{\rm endo}_{\mathbb{C}}
(\mathbb{C}^n)$.  This implies the matrices $A$ and $B$ must satisfy the relations: 
$a A A^\dagger + b B B^\dagger = aI_n$ and $a B B^\dagger+b A A^\dagger=b I_n$ and 
hence $ab(a^2-b^2) B B^\dagger=0$.  As $0<a,b,<1$ and $a\ne b$ by assumption, we 
must have $B=0$ and $A\in\uu(n)$. Thus, the isotropy subgroup at $\rho$ is $\left(
\begin{array}{cc} A & 0 \\ 0 & A^*\end{array}\right)$ with $A\in \uu(n)$.  By Theorem
\ref{thm:one}, the orbit of $\rho$ is a homogeneous manifold of real dimension 
$n(2n+1)-n^2=n^2+n$.
\end{proof}

\begin{proposition} \label{prop:sp:four}
If $\rho=\diag(D_{2n-2\ell},\alpha I_{2\ell})$, where $D_{2n-2\ell}$ is a diagonal 
$(2n-2\ell)\times (2n-2\ell)$ complex matrix, $I_{2\ell}$ is the identity $2\ell
\times 2\ell$ complex matrix, and $\alpha\in [0,1]$, then the orbit of $\rho$ is a
homogeneous manifold of real dimension at most $n(2n+1)-\ell(2\ell+1)$.
\end{proposition}

\begin{proof}
We note the isotropy subgroup at $\rho$ contains the group 
$\left(\begin{array}{cc} I_{2n-2\ell}&0 \\ 0 & Q_{2\ell}\end{array}\right) \ |\ 
Q\in \sp(\ell)\}$, which is isomorphic to the subgroup $\sp(\ell)$.  Hence, the 
orbit of $\rho$ is a homogeneous manifold of real dimension at most $n(2n+1)-
\ell(2\ell+1)$ as desired.
\end{proof}

In table \ref{table:two} we compare the dimensions of the orbits of various types 
of ensembles under $\uu(2n)$ with those of the orbits under $\sp(n)$ for $n=2$ and 
$n=3$.  The classification of the orbits is based on the spectrum of $\rho$.  Note
that the spectrum of $\rho$ uniquely determines the unitary orbit (or equivalence
class) $\rho$ belongs to.  However, two density matrices with the same spectrum may
belong to different symplectic orbits.  For instance, $\rho_0=\diag(a,b,a,b)$ and 
$\rho_1=\diag(a,a,b,b)$ are unitarily equivalent but belong to different orbits 
under the symplectic group as defined above.  See example 1 in \cite{JPA35p8551}.

\begin{table}
\[\begin{array}{|l|l|l|}
\hline
N=4                   & \uu(4)  & \sp(2)  \\ \hline
\rho = \diag(a,a,a,a) & 0       & 0 \\
\rho = \diag(a,b,b,b) & 6       & 6 \\
\rho = \diag(a,a,b,b) & 8       & 6 \\
\rho = \diag(a,b,c,c) & 10      & \le 8 \\
\rho = \diag(a,b,c,d) & 12      & \le 8 \\
\hline
\end{array}\]
\[\begin{array}{|l|l|l|}
\hline
N=6                   & \uu(6) & \sp(3) \\ \hline
\rho=\diag(a,a,a,a,a,a) & 0   & 0 \\
\rho=\diag(a,b,b,b,b,b) & 10  & 10 \\
\rho=\diag(a,a,b,b,b,b) & 16  & \le 21-10=11 \\
\rho=\diag(a,b,c,c,c,c) & 18  & \le 21-10=11 \\
\rho=\diag(a,b,b,c,c,c) & 22  & \le 2\times 3^2=18  \\
\rho=\diag(a,b,c,d,d,d) & 24  & \le 18  \\
\rho=\diag(a,a,b,b,c,c) & 24  & \le 18  \\
\rho=\diag(a,a,a,b,b,b) & 18  & \le 3^2+3 = 12 \\
\rho=\diag(a,b,c,c,d,d) & 26  & \le 18  \\
\rho=\diag(a,b,c,d,e,e) & 28  & \le 18  \\
\rho=\diag(a,b,c,d,e,f) & 30  & \le 18 \\
\hline
\end{array}\]
\caption{Dimensions of the orbits of various types of ensembles classified by their
spectrum under the unitary group $\uu(2n)$ and the symplectic group $\sp(n)$ for 
$n=2$ and $n=3$.  The table shows that the orbits under the symplectic group are
in general much smaller than the orbits under the unitary group, except in the 
case of pure-state-like (and completely random) ensembles, for which the sympletic
orbits have the same dimension as the unitary orbit.  Note that in all cases it is
assumed that different letters $a,b,\ldots$ represent different values in $[0,1]$.}
\label{table:two}
\end{table}

\section{Conclusion}

We have shown that the action of a Lie group on the set of quantum states endows
it with the structure of a stratified set with strata given by the orbit manifolds.
In particular, we studied the stratification of the set of states induced by the 
action of the unitary group, which is especially useful since the unitary orbits
are of interest in quantum control and computing, where they determine the maximal
set of quantum states that are reachable from a given set via a coherent control 
or by applying a unitary gate.  Furthermore, there are many properties of quantum
states such as von-Neumann or Renyi entropy that depend only on the unitary orbit
the state belongs to.  It therefore makes sense to define these functions on the
unitary stratification.  We have shown that the unitary orbits can be identified
with flag manifolds whose type and dimension depend only on the \emph{multiplicity}
of the eigenvalues of the states belonging to the orbit.  We have also determined
the dimensions of the orbit manifolds and shown that we can define a partial 
ordering related to the degree of disorder in the system on this stratification
via majorization.

To better understand the geometry and structure of the set of quantum states, we
studied the embedding of the quantum states and their associated orbit manifolds
of an $n$-level system into real Euclidean space provided by the coherence vector.
We showed that the coherence vector we defined always maps the quantum states into
a closed ball in $\mathbb{R}^{n^2-1}$ in such a manner that the orbit manifolds 
are mapped to submanifolds of spheres of fixed radius from the center.  For $n=2$ 
this embedding is surjective, hence justifying the identification of the set of 
quantum states with the closed ball in $\mathbb{R}^3$, and the identification of
the orbits with concentric spheres inside this ball.  By comparing the dimensions
of the orbits we also showed that this embedding is \emph{no} longer surjective 
for $n>2$ and the orbit manifolds in this case are proper submanifolds of spheres
in $\mathbb{R}^{n^2-1}$ of lower dimension.  The manifold of pure states is always
a submanifold of the boundary sphere, which contains no other orbits, while all 
other spheres of fixed distance from the center generally contain infinitely many
disjoint orbits of varying dimensions, and depending on the distance of the sphere
from the center, a positive measure set of points which do not correspond to 
quantum states at all.  A detailed analysis for the three-level case was provided.

Finally, we studied systems whose natural evolution is restricted to a subgroup 
of the unitary group such as the symplectic group due to dynamical symmetries. 
We showed that we can define a refined stratification based on the smaller orbits
of this subgroup.  In case of the symplectic group we have shown that the orbits
of all pseudo-pure states agree with the unitary orbits, while the symplectic 
orbits of all other (mixed) states have lower dimension than the unitary orbits.  
From the point of view of control this means that we can control pure and pseudo
pure states for such systems but not generic ensembles.

\ack
SGS would like to thank A.~I.\ Solomon (The Open University) and D.~K.~L.\ Oi
and A.~K.\ Ekert (University of Cambridge) for helpful discussions and for 
reading early drafts of the manuscript and providing numerous comments and 
suggestions that have improved it substantially, as well as the Cambridge-MIT 
Institute for financial support.

\appendix
\section{The symplectic group} \label{appendix:A}

The symplectic group $\sp(n)$ is usually defined as the Lie group of automorphisms
on $\mathbb{H}^n$, where $\mathbb{H}$ is the skew-field of quaternions, that 
preserve the canonical symplectic inner product
\[
  \ip{\vec{q}}{\vec{q'}} = \sum_{i=1}^n \bar{q}_i q_i',
\]
where $\vec{q}$ and $\vec{q}'$ are $n$-vectors whose entries $q_i$ are quaternions
and conjugation $\bar{\vec{q}}$ is over $\mathbb{H}$.

The (skew-field) of quaternions  $\mathbb{H}$ can be regarded as a vector space 
over $\mathbb{R}$ with the standard basis $\{\vec{1},\vec{e}_1,\vec{e}_2,\vec{e}_3\}$ 
subject to the multiplicative relations: $\vec{e}_i^2=-1$ for $1\le i\le 3$ and 
$\vec{e}_i\vec{e}_j=-\vec{e}_j\vec{e}_i=\vec{e}_k$ for any even permutation $(i,j,k)$
of the set $(1,2,3)$.  Since the field of complex numbers $\mathbb{C}$ is isomorphic
to $\mathbb{R}\cdot \vec{1} \oplus \mathbb{R}\cdot \vec{e}_1$ and every quaternion $q$
can be written as 
\[
  q = q_0 + q_1 \vec{e}_1 + q_2 \vec{e}_2 + q_3 \vec{e}_3 
    = (q_0 + q_1\vec{e}_1) + \vec{e}_2(q_2-q_3\vec{e}_1),
\]
we may also regard $\mathbb{H}$ as a vector space over $\mathbb{C}$ with basis
$\{\vec{1}, \vec{e}_2\}$.  We therefore obtain an isomorphism $\phi:\mathbb{H}^n
\rightarrow \mathbb{C}^{2n}$ of complex vector spaces via $\phi(q_1,\cdots,q_n)=
(z_1,\cdots, z_{2n})$ where $q_i=z_i+ z_{n+i}\vec{e}_2$.  Consequently, 
\[
  \ip{\vec{q}}{\vec{q}'} 
  = \sum_{i=1}^{n} \bar{\vec{q}}_i \vec{q}_i'
  = \left(\sum_{i=1}^{2n} z_i^* z'_i \right) + 
      \vec{e}_2 \left(\sum_{i=1}^n (z_i z_{n+i}'-z_{n+i}z_i') \right). 
\]

The isomorphism $\phi$ allows us to identify an isometry $A$ of $\mathbb{H}^n$, 
with a complex automorphism of $\mathbb{C}^{2n}$ that preserves both the canonical 
Hermitian inner product and the canonical skew-symmetric bilinear form on 
$\mathbb{C}^{2n}$ defined by 
\[ 
  S(\vec{z},\vec{z}') = \sum_{i=1}^n (z_i z'_{n+i}-z_{n+i}z_i'),
\]
where $\vec{z}=(z_1,\cdots, z_{2n})$ and $\vec{z}'=(z_1',\cdots, z_{2n}')$.  

Since $A$ preserves the canonical Hermitian inner product on $\mathbb{C}^{2n}$, 
$A\in\uu(2n)$.  Since $A$ leaves invariant the canonical skew-symmetric bilinear
form $S(\cdot,\cdot)$, it is equivalent to having $A^T J A=J$. 
\\




\begin{thebibliography}{10}

\bibitem{Bloch} 
F. Bloch, Phys. Rev. {\bf 70}, 460 (1946)

\bibitem{Ruskai} 
M. B. Ruskai, S. Szarek and E. Werner,
Linear algebra and its applications {\bf 347}, 159-187 (2002);
arXiv: quant-ph/0101003.

\bibitem{Hioe} F. T. Hioe and J. H. Eberly, 
Phys. Rev. Lett. {\bf 47}, 838 (1981)

\bibitem{Lendi} K. Lendi, 
Phys. Rev. A {\bf 34}, 662 (1986)

\bibitem{Alicki} R. Alicki and K. Lendi, 
\textit{Quantum Dynamical Semigroups and Application},
Lecture Notes in Physics Vol. 286, Springer Verlag (Berlin, 1987)

\bibitem{Kimura} G. Kimura, 
\textit{The Bloch Vector for $N$-level Systems}
Phys. Lett. A {\bf 314}, 339 (2003) 

\bibitem{Byrd} M. S. Byrd and N. Khaneja, 
arXiv: quant-ph/0302024 v2 (2003)

\bibitem{Nielson}
M. A. Nielson, 
\textit{Characterizing mixing and measurement in quantum mechanics}
arXiv: quant-ph/0008073 (2000)

\bibitem{refGB}  G. Bredon, \textit{Topology and Geometry},
Graduate Texts in Mathematics 139, Springer-Verlag (Berlin, 1993)

\bibitem{JPA35p4125}
S. G. Schirmer and J. V. Leahy and A. I. Solomon,
J. Phys. A {\bf 35}, 4125 (2002)

\bibitem{JPA35p8551}
S. G. Schirmer and A. I. Solomon and J. V. Leahy,
J. Phys. A {\bf 35}, 8551 (2002)

\end{thebibliography}
\end{document}